# Title: Enhanced Kerr electro-optic nonlinearity through cascaded Pockels effects

**Authors:** Guang-Zhen Li, Yu-Ping Chen, Hao-Wei Jiang, and Xian-Feng Chen

**Affiliations:**

[1]State Key Laboratory of Advanced Optical Communication Systems and Networks, Department of Physics and Astronomy, Shanghai Jiao Tong University, 800 Dongchuan Road, Shanghai 200240, China

[2]Key Laboratory for Laser Plasma (Ministry of Education), IFSA Collaborative Innovation Center, Shanghai Jiao Tong University, Shanghai 200240, China

**The email addresses of all authors:**

liguangzhen520@sjtu.edu.cn, ypchen@sjtu.edu.cn, jsjdjhw@sjtu.edu.cn, xfchen@sjtu.edu.cn

**The corresponding authors:**

YP Chen, Department of Physics and Astronomy, Shanghai Jiao Tong University, 800 Dongchuan Road, Shanghai, China 200240

Email: ypchen@sjtu.edu.cn     Tel: +86-21-54743273     Fax: +86-21-54747249

XF Chen, Department of Physics and Astronomy, Shanghai Jiao Tong University, 800 Dongchuan Road, Shanghai, China 200240

Email: xfchen@sjtu.edu.cn     Tel: +86-21-54743252    Fax:+86-21-54747249

## ABSTRACT

We demonstrated a large enhancement of Kerr electro-optic nonlinearity through cascaded Pockels effects in a domain inversion ferroelectric crystal. We designed a structure that can implement the cascaded Pockels effects and second-harmonic generation simultaneously. The energy coupling between the fundamental lights of different polarizations led to a large nonlinear phase shift, and thus an effective electro-optic nonlinear refractive index. The effective nonlinearity can be either positive or negative, causing the second-harmonic spectra to move towards the coupling center, which in turn, offered us a way to measure the effective electro-optic nonlinear refractive index. The corresponding enhanced Kerr electro-optic nonlinearity is more than three orders of magnitude higher than the intrinsic value. These results open a door to manipulate the nonlinear phase by applying external electric field instead of light intensity in noncentrosymmetric crystals.

## INTRODUCTION

The third-order nonlinearities, though with weak third-order coefficients, [1,2] exist in a medium with any symmetry.[3-5] One common way to enhance the intrinsic weak third-order nonlinearity is via cascading second order nonlinear effects, [6] because of its much higher value than the direct higher-order nonlinearity. [7-9] Kerr electro-optic (EO) effect is connected to the appearance of the nonlinear third-order susceptibility. [1] It takes the advantage of the modulation of the electric field and intrinsic nature of fast response time. Besides applications in electro-optic switching, [10] electro-optical detection, [11] high-speed optical shutters, [12] it is also used to measure the optical third-order susceptibility of material. [13] However, Kerr EO effect is relatively weak in noncentrosymmetric crystals for the existence of linear EO effect. [1] Therefore, it is highly demanded to enhance the Kerr EO nonlinearity and broaden its applications in noncentrosymmetric crystals.

In this paper, we demonstrated a scheme of enhancing Kerr EO nonlinearity through cascaded linear EO effects ($\gamma:\gamma$) working near its phase-matching condition in a domain inversion ferroelectric crystal-MgO doped periodically poled lithium niobate (PPLN). The induced nonlinear phase shift gave rise to an effective EO nonlinear refractive index, corresponding a large effective Kerr EO coefficient. This effective nonlinearity can be either positive or negative, depending on the sign of wave-vector mismatching during the cascading processes. The enhanced Kerr EO nonlinearity is more than three orders of magnitude higher than the intrinsic value. We also found that the cascaded linear EO effects and second-harmonic generation (SHG) could be implemented simultaneously for a given inversion domain period, as long as choosing a properly operating fundamental wave (FM) and experimental temperature. Consequently, the enhanced Kerr EO nonlinearity can also control the process of SHG.

## MATERIALS AND METHODS

### Principle of cascaded linear EO effects

When an external electric field is applied along the y-axis of a LiNbO$_3$ crystal, [1] the principle axes of the new index ellipsoid rotates with an angle of $\theta \approx (\gamma_{51}E_y - s_{41}E_y^2)/[1/(n_o^\omega)^2 - 1/(n_e^\omega)^2]$ with respect to the unperturbed principle axes. Taking both linear and Kerr EO effects into account, we deduce the refractive index of the new optical axis due to the equation of the index ellipsoid,

$$n_z' = n_e^\omega - \frac{1}{2}s_{13}(n_e^\omega)^3 E_y^2 + \frac{1}{2}(\gamma_{51}E_y - s_{41}E_y^2)(n_e^\omega)^3 \tan\theta. \qquad (1)$$

$n_e^\omega$ and $n_o^\omega$ represent the indices of the fundamental extraordinary and ordinary waves, $\gamma_{51}$ and $s_{13}$, $s_{41}$ are the linear and quadratic electro-optic coefficients, and $E_y$ is the y-axis external electric field, respectively. As for periodically poled LiNbO$_3$ crystal (Figure 1b and 1e), the c axis of the LiNbO$_3$ is inverted periodically. It leads to the periodic alteration of the sign of nonlinear optical susceptibility and electro-optic coefficients. Therefore, when the electric filed is applied along the y-axis of PPLN, optical axis of each domain rotates periodically as shown in Figure 1a. [1,14] Then the energy of the incident e-polarized wave will flow to the generated o-polarized wave and then it will flow back. If it occurs near its phase-matching condition, the returning e-polarized wave will have a different phase from the original e-polarized wave that does not deplete completely as dedicated in Figure 1b-1c.

The amplitude of e-polarized wave is solved by the coupled-mode equations, [1] which is $A(z) = e^{-i(\Delta\beta/2)z}[\cos(sz) + i\Delta\beta\sin(sz)/(2s)]$. $\Delta\beta = 2\pi(n_o^\omega - n_e^\omega)/\lambda - 2\pi/\Lambda$ is the wave-vector mismatching for cascaded linear EO effects, $s = [\kappa\kappa^* + (\Delta\beta/2)^2]^{1/2}$, $\kappa$ (ref.15) is the coupled coefficient, $\Lambda$ is the domain period of PPLN and $\lambda$ is the fundamental wavelength, respectively. Then we obtain the nonlinear phase change impressed onto the fundamental e-polarized wave at the exit surface $z = L$, which is

$$\Delta\Phi_e^{NL} = \frac{\Delta\beta L}{2} - \arctan[\frac{\Delta\beta}{2s}\tan(sL)]. \quad (2)$$

L is the length of the crystal. We can also achieve an EO effective nonlinear refractive index $\Delta n_2^{eff}$ deduced by $\Delta\Phi_e^{NL}$, since $\Delta\Phi_e^{NL} = (2\pi L/\lambda)\Delta n_2^{eff}$ (ref.16). We plot the calculated transmission spectrum [17] and the effective EO nonlinear refractive index as a function of $\Delta\beta L$ in Figure 2. The effective nonlinearity can be either positive or negative, depending on the sign of $\Delta\beta$. We can see that $\Delta\Phi_e^{NL} > 0$, $\Delta n_2^{eff} > 0$ for $\Delta\beta < 0$ and $\Delta\Phi_e^{NL} < 0$, $\Delta n_2^{eff} < 0$ for $\Delta\beta > 0$. It is similar to the phase shift caused by cascaded $\chi^{(2)}:\chi^{(2)}$ process. [8,18] $\Delta\beta = 0$ corresponds to the central fundamental wavelength $\lambda_c$ of the transmission spectrum. $\Delta\beta < 0$ refers the region that the fundamental wavelength $\lambda > \lambda_c$, and $\Delta\beta > 0$ refers the region that $\lambda < \lambda_c$.

In the limit of weak cascaded effects and negligible depletion of the fundamental wave, the nonlinear phase shift is approximately proportional to the square of the electric field $E_y$. And the effective EO nonlinear refractive index is deduced by

$$\Delta n_2^{eff} \approx -\frac{2(n_o^\omega n_e^\omega)^3 \gamma_{51}^2}{\pi\lambda\Delta\beta} E_y^2. \quad (3)$$

Here $|\Delta\beta| \gg |\kappa|$ should be satisfied under large phase mismatching or a low external electric field. In this case, $\Delta n_2^{eff}$ is approximately proportional to the square of the external electric field and independent of the incident optical intensity. Otherwise the approximation breaks down, Eq. (3) must be solved exactly.

Thus the index variations induced by different EO effects should be expressed by $\Delta n = \Delta n_1 + \Delta n_2 + \Delta n_2^{eff}$ as shown in Figure 1d. $\Delta n_1$ and $\Delta n_2$ are the changes of refractive indices by the linear and intrinsic Kerr EO effects, whose relationships with electric fields are plotted in Figure 4a. The EO coefficients of LiNbO$_3$ are $\gamma_{51} = 32.6 \times 10^{-12}$ m/V (ref.1) and $s_{13} = 2.3 \times 10^{-21}$ m$^2$/V$^2$ (ref.19), respectively. Supposing the external electric field is 0.1 V/μm (@ $\lambda$ = 1581.9 nm, $\Delta\beta > 0$), the magnitude of the rotation angle $\theta$ is $10^{-4}$. Therefore we can get $\Delta n_1 = 1.6 \times 10^{-9}$, $\Delta n_2 = 1.1 \times 10^{-10}$ and $\Delta n_2^{eff} = -1.1 \times 10^{-6}$, which are marked as point A, B, C in Figure 4, respectively. $\Delta n_1$ and $\Delta n_2$ can be ignored if compared with $\Delta n_2^{eff}$, which makes all the index changes $\Delta n \approx \Delta n_2^{eff}$.

Just like $\chi^{(2)} : \chi^{(2)}$ inducing an effective $\chi^{(3)}_{eff}$ in cascaded second-order nonlinearities,[8] $\gamma_{ij} : \gamma_{ij}$ leads to an effective Kerr EO coefficient $s_{ij}^{eff}$ in cascaded linear EO effects. Then the refractive index modified by the cascaded effects can be expressed as

$$n = n_e^\omega + \frac{1}{2}\gamma_{ij}(n_e^\omega)^3 E_y + \frac{1}{2}s_{ij}^{eff}(n_e^\omega)^3 E_y^2 + \frac{1}{2}s_{ij}(n_e^\omega)^3 E_y^2 + \cdots. \quad (4)$$

We get $s_{13}^{eff} = -2\Delta n_2^{eff} /[(n_e^\omega)^3 E_y^2] \approx 4(n_o^\omega)^3 \gamma_{51}^2 /(\pi\lambda\Delta\beta)$ from Eq.(1), (2) and (4). The calculated value of the effective Kerr EO coefficient is $s_{13}^{eff} \approx 2.2 \times 10^{-17}$ m$^2$V$^{-2}$. It is more than three orders of magnitude higher than the intrinsic value.[19]

**Manipulation of SHG**

For quasi-phase-matching (QPM) SHG,[20,21] the wave vector mismatching is given by $\Delta k = 4\pi(n_e^{2\omega} - n_e^\omega)/\lambda - 2\pi/\Lambda'$ with the domain period of $\Lambda'$. $n_e^{2\omega}$ is the index of second-harmonic extraordinary wave. Refractive indices are calculated by Sellmeier equations.[22] Supposing $\Lambda = \Lambda'$, the two processes can be realized in a single PPLN with a proper incident wavelength and temperature simultaneously. The induced EO nonlinear refractive index affects the original wave-vector mismatching of SHG effectively, which makes it turn into

$$\Delta k' = \frac{4\pi}{\lambda}[n_e^{2\omega} - (n_e^\omega + \Delta n_2^{eff})] - \frac{2\pi}{\Lambda'} \quad (5)$$

**Measurement of transmission and SHG spectra**

Figure 1e shows one part of etched poling surface of the z-cut 5% MgO doped periodically poled LiNbO3 crystal, with domain inversion period of 20.3 μm and a dimension of 40 × 10 × 0.5 mm . The external electric field is applied along the y-axis of PPLN, and light propagates along x-axis. The light from the tunable continuous laser (1517-1628 nm) was amplified to 100 mW (corresponding to 1.6×10$^5$ W/cm$^2$) by erbium-doped fiber amplifier.

The sample was placed between two parallel polarization beam splitters, which constituted the Solc filters.[17] A high voltage source with the maximum of 10 kV was used to generate the external electric field along the y-axis of the PPLN. One power meter working on c-band measured the transmission of the output fundamental wave, while another power meter working on visible light measured the intensity of second-harmonic wave, respectively.

## RESULTS AND DISCUSSION

We observed that the two processes, namely, the cascaded linear EO effects and SHG, occurred simultaneously at the wavelength of 1582.1 nm and the same temperature of 26.3 °C. The overlapped spectra are plotted in Figure 3a. When the experimental temperature is changed, the two spectra separated from each other at an opposite direction. We measured the intensity of SHG for different external electric fields at a fixed experimental temperature. Figure 3b-3d shows the results at different temperatures. The shift of the SHG spectra is due to the variable effective EO nonlinear refractive index induced by different electric fields.

As seen in Figure 3a-3b, at 26.3 °C, the absolute value $|\Delta k'|$ became larger at both sides of SHG spectrum along with the increase of the applied electric field, because the spectra are fully overlapped. It led to a dramatic decrease of the SHG efficiency. At 24.1 °C as shown in Figure 3c, SHG spectrum is located at left region of transmission gap ($\lambda < \lambda_c, \Delta\beta > 0$), in which $\Delta n_2^{eff} < 0$. $|\Delta k'|$ became larger at left side and smaller at the other side of SHG spectrum. As a result, the whole SHG spectrum moved right. Oppositely, at 27.6 °C in Figure 3d, SHG spectrum located at the right region of transmission gap ($\lambda > \lambda_c, \Delta\beta < 0$). The positive $\Delta n_2^{eff}$ led to the SHG spectrum shifting left.

Figure 3 suggests how the enhanced Kerr EO nonlinearity controls the process of SHG. On the basis of which, we can measure the magnitude of $\Delta n_2^{eff}$ according to the shift of the SHG central wavelength. The experimental results calculated from Figure 3c-3d are plotted in Figure 4b, where $\Delta n_2^{eff} < 0$ in Figure 3c and $\Delta n_2^{eff} > 0$ in Figure 3d, respectively. They satisfy the condition of large phase-mismatching and the variations of the effective nonlinear indices are proportional to the square of the external electric field. It is in good agreement with the simulation results deduced by Eq. (3). At $E_y = 0.1 \, \text{V/μm}$ (@1581.9 nm, $\Delta\beta > 0$), the experimental values are $\Delta n_2^{eff} = -6.4 \times 10^{-7}$, and we get $s_{13}^{eff} \approx 1.3 \times 10^{-17} \, \text{m}^2\text{V}^{-2}$, which are identical to the theoretical values.

In general, there are three other possible effects that may contribute to the shift of the SHG spectra, including the intensity of the fundamental light, the index change caused by intrinsic EO effects, and the cascaded nonlinearity between the second-harmonic and fundamental wave. However, in our scheme, all of them are not possible to came into play. First, since the intensity of SHG is proportional to the square of the intensity of fundamental wave,[2] we observed the normalized transmission and SHG intensity as a function of the applied electric field. As demonstrated in Figure 5, we selected two wavelengths of 1581.8 nm and 1582.3 nm from each sideband of SHG spectrum in Figure 3c. At $E_y = 0.32$ V/μm, the same normalized transmittances were measured for these two different wavelengths. However, for their normalized SHG intensities, the one is high to 0.81 at A and the other is low to 0.29 at B. It means that although the intensity of incident light varies with the external electric field, it hardly affects the efficiency of SHG. Second, when the temperature was high enough that the transmission spectrum is barely overlapped with SHG, we observed that the SHG spectrum remained unchanged when varying the external electric fields. It agrees well with our discussion that the index variation caused by intrinsic EO effects are small so it can be neglected. Last, the cascaded second-order nonlinear process is invalid in this case because the incident optical intensity is pretty low (1.6×10$^5$ W/cm$^2$).

We also observed the same phenomena at domain periods of 20.1 μm and 19.9 μm, which makes it significant to explore the further intrinsic bond between the two physical processes. The wave-vector mismatchings, $\Delta\beta$ and $\Delta k$ are determined by the dispersion relations,[22] as a function of the wavelength, domain inversion period and temperature. Supposing $\Delta\beta$ and $\Delta k$ equal to zero simultaneously, the relationship between the domain period and wavelength is plotted in Figure 6. Points a, b and c correspond to the three inversion domain periods we performed in our experiment. The inaccuracy of Sellmeier equations causes the theoretical wavelength (1583.1 nm) for 20.3 μm a little shift from the experimental condition (1582.1 nm). By careful calculation, we confirm that these two processes can be satisfied simultaneously at a designed domain inversion period if employing a proper wavelength and temperature. The corresponding relationship is inserted in Figure 6. Therefore if given one of the three parameters in Figure 6, we can find the other two, which is significant in practical flexibility and adjustability.

## CONCLUSIONS

In conclusion, we observed a large effective EO nonlinear refractive index and an enhancement of Kerr EO nonlinearity through cascaded linear EO effects when an external electric field was applied in PPLN. The enhanced Kerr EO nonlinearity is more than three orders of magnitude higher than the intrinsic value. Moreover, besides SHG, other second-order parametric processes such as sum and different frequency generation, can also be manipulated by this enhanced Kerr EO nonlinearity. The principle basis of this Kerr EO nonlinearity is quite different from that induced by the cascaded second-order nonlinear processes, for its independence of the light intensity. Therefore, it can find potential applications in electrically controlled third-order nonlinearities, such as group velocity control, phase modulation, etc..

## ACKNOWLEGEMENTS

The research was supported by the National Natural Science Foundation of China under Grant No.11174204, 61125503, 61235009, the Foundation for Development of Science and Technology of Shanghai under Grant No. 13JC1408300.

## FIGURES

**Figure 1**   (a) The optical axes of positive and negative domains of PPLN rotate by angle of $\theta$ and $-\theta$ under y-axis external electric field, respectively. (b) Schematic of achieving cascaded linear EO effects and SHG simultaneously with an external electric field being applied onto the PPLN along y-axis. The periodically inverted optical axes of PPLN lead to the periodic alteration of the sign of electro-optic coefficients ($\pm\gamma_{ij}$). (c) Illustration of cascaded linear EO effects. The energy flows from the incident e-polarized wave to the regenerated o-polarized wave and then flows back near its phase-matching condition ($\Delta\beta$), inducing a nonlinear phase shift. (d) Changes of refractive indices caused by linear ($\gamma_{ij}$), Kerr ($s_{ij}$), and cascaded linear ($\gamma_{ij}:\gamma_{ij}$) electro-optic effects, respectively. (e) A part of etched poling surface of the sample, with domain inversion period of 20.3 μm.

**Figure 2** The calculated transmission spectrum (red) and the effective EO nonlinear refractive index (blue) as a function of $\Delta\beta L$. $\Delta\beta = 0$ corresponds to the central wavelength $\lambda_c$ of the transmission spectrum. $\Delta\beta < 0$ refers to the region that the fundamental wavelength $\lambda > \lambda_c$, where $\Delta n_2^{eff} > 0$. $\Delta\beta > 0$ refers to the region $\lambda < \lambda_c$, where $\Delta n_2^{eff} < 0$.

**Figure 3** (a) Measured transmission (red) and SHG spectra (black) that are fully overlapped at central fundamental wavelength of 1582.1 nm, at T=26.3 °C. (b-d) SHG spectra with varied external electric fields. (b) At 26.3 °C, the spectra fully overlapped. The wave-vector of SHG, $|\Delta k'|$ became larger at both sides of SHG spectrum, which decreased the efficiency of SHG. (c) At 24.1 °C, SHG spectrum located at the left region of transmission spectrum ($\lambda < \lambda_c, \Delta\beta > 0$), where $\Delta n_2^{eff} < 0$ led the SHG spectrum shift right. (d) At 27.6 °C, SHG spectrum located at the right region of transmission gap ($\lambda > \lambda_c, \Delta\beta < 0$). The positive $\Delta n_2^{eff}$ led the SHG spectrum shift left.

**Figure 4** (a) The index variations $\Delta n_1$ (dashed) and $\Delta n_2$ (solid), caused by the linear and intrinsic Kerr EO effects as a function of the external electric field. (b) Nonlinear refractive index $\Delta n_2^{eff}$ caused by cascaded linear EO effects versus the external electric field. For the specific case of $\lambda = 1581.9$ nm ($\Delta\beta > 0$, $\Delta n_2^{eff} < 0$) in Figure 3c and $\lambda = 1582.6$ nm ($\Delta\beta < 0$, $\Delta n_2^{eff} > 0$) in Figure 3d, they satisfy the condition of large phase-mismatching. The experimental (dots) results are in good agreement with the simulation (solid line) that $\Delta n_2^{eff}$ is proportional to the square of the external electric field. Points A, B and C mark the index changes at $E_y = 0.1$ V/μm. $\Delta n_1$ and $\Delta n_2$ can be ignored as they are much smaller than $\Delta n_2^{eff}$, and thus $\Delta n \approx \Delta n_2^{eff}$.

**Figure 5** Measured normalized transmission (a) and SHG intensity (b) at two selected wavelengths (1581.8 nm and 1582.3 nm in Figure 3c) as a function of the external electric fields. At $E_y = 0.32$ V/μm, the two wavelengths have the same transmittances, but quite different SHG intensities. It means that the intensity of FW hardly affects the efficiency of SHG and the observed shift of SHG spectra is caused by the cascading effects.

**Figure 6** Calculated inversion domain periods for achieving SHG (solid) and the cascading Pockels effects(dashed) as a function of fundamental wavelengths at different temperatures, calculated by Sellmeier functions. Points a, b and c correspond to the three inversion domain periods we performed in our experiment. The inset figure shows the relationship among the three parameters to realize the cascading process and SHG simultaneously.

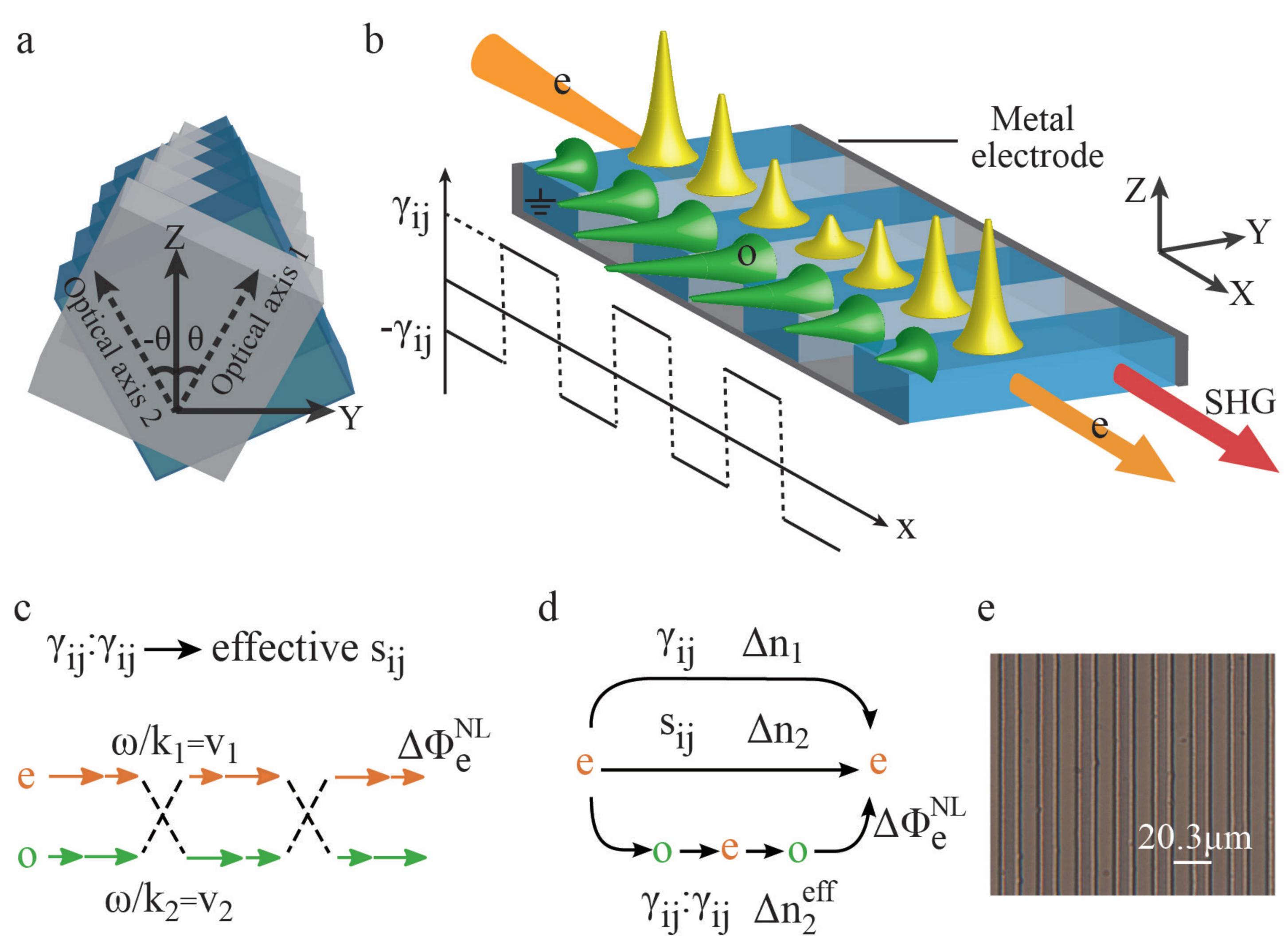

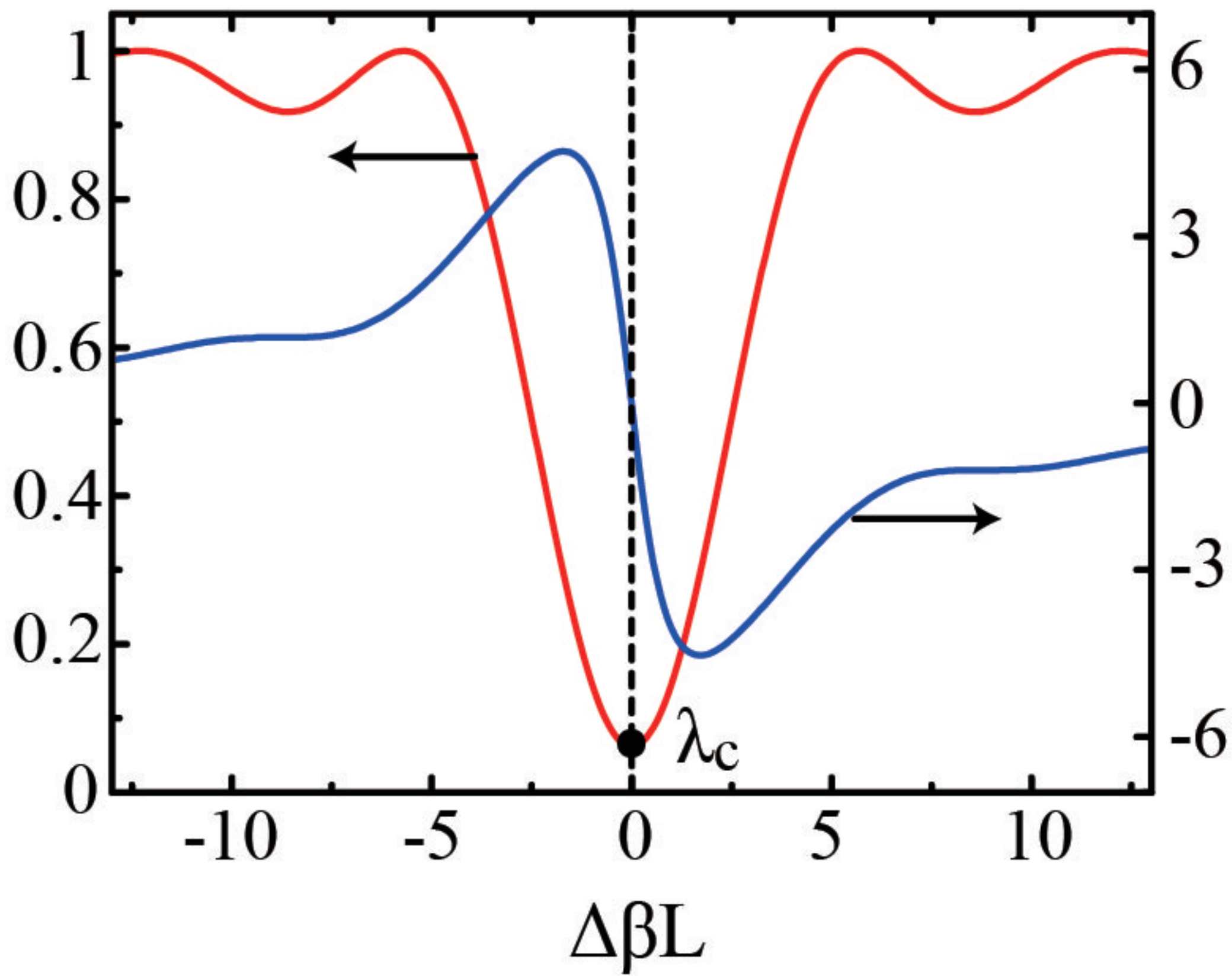

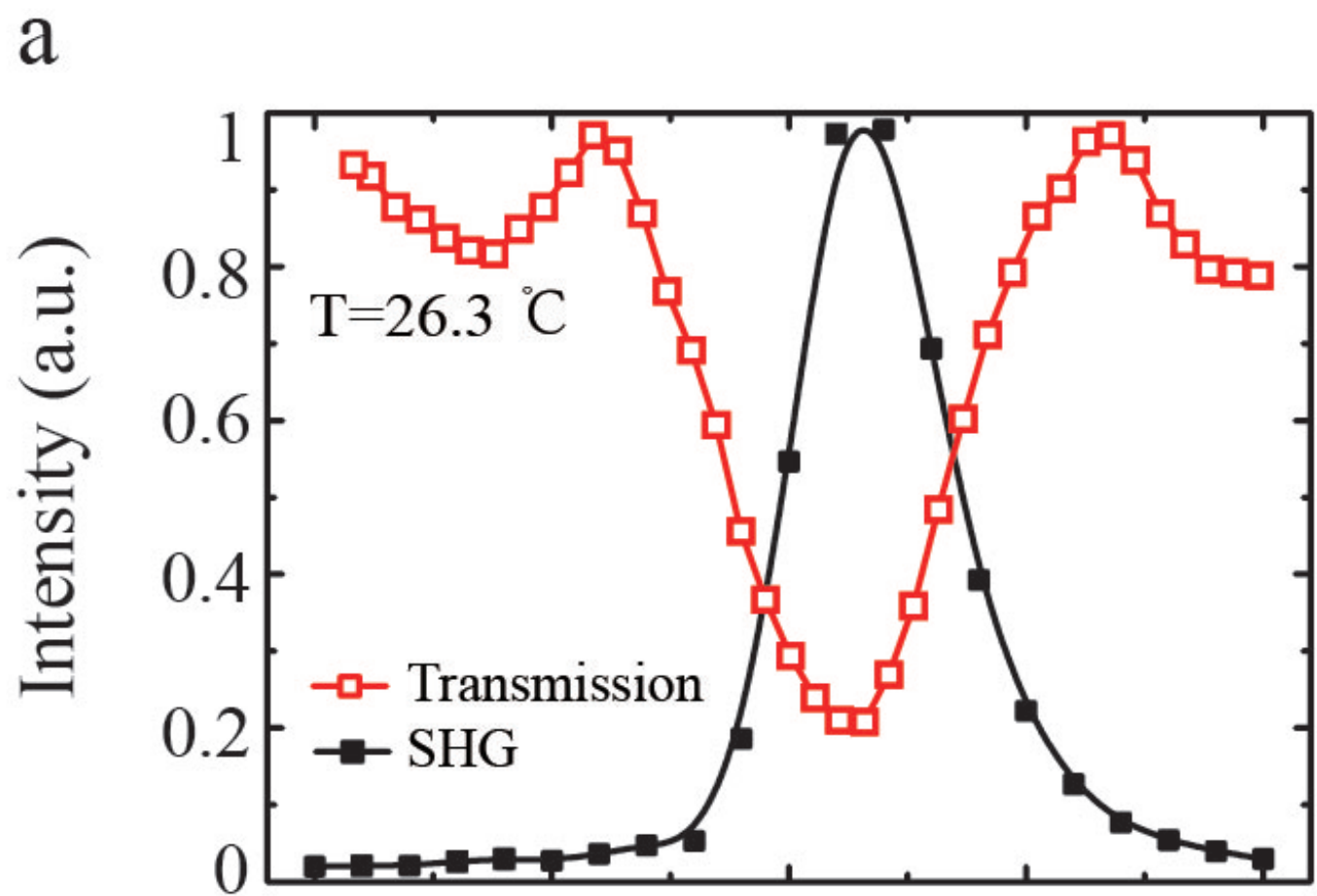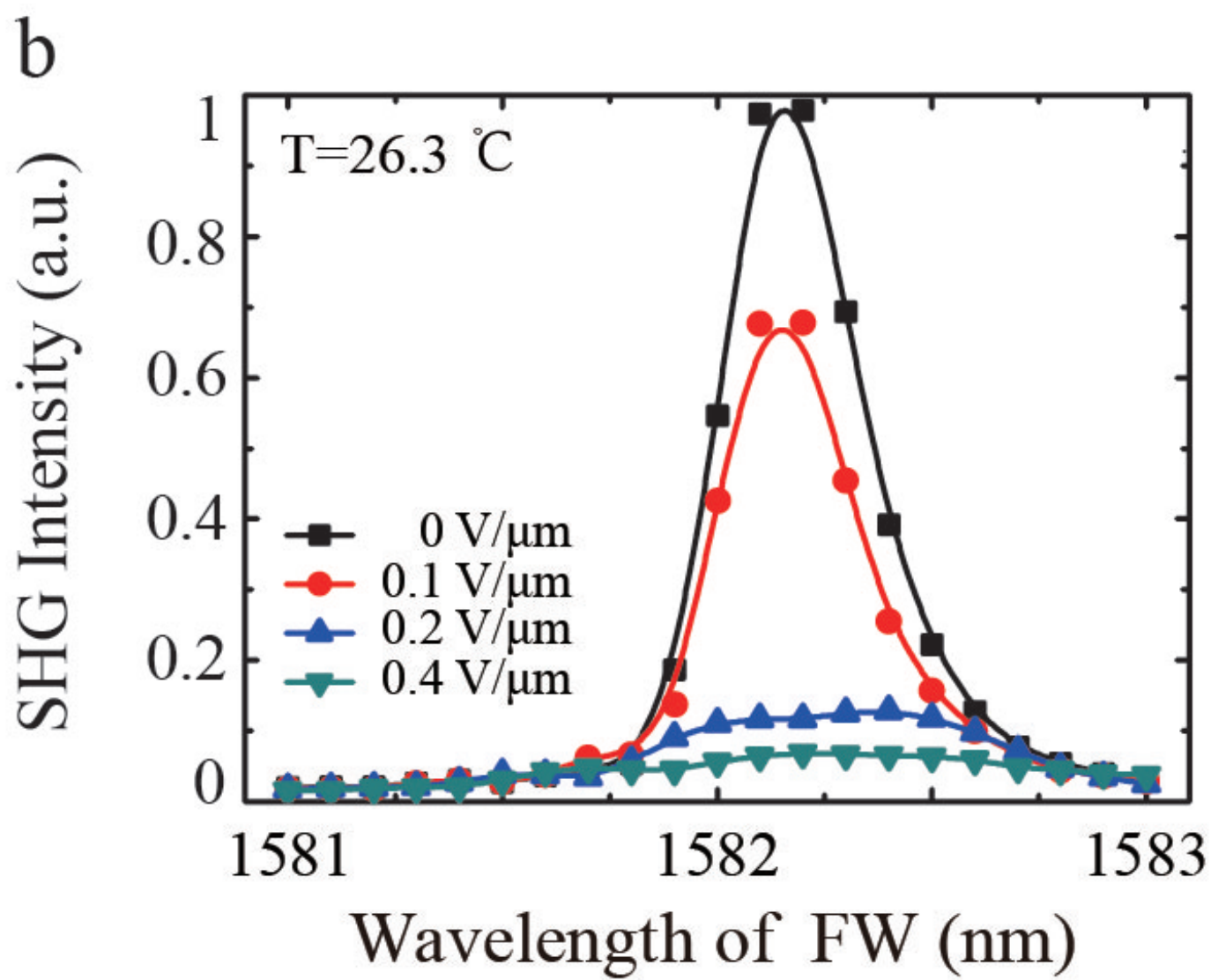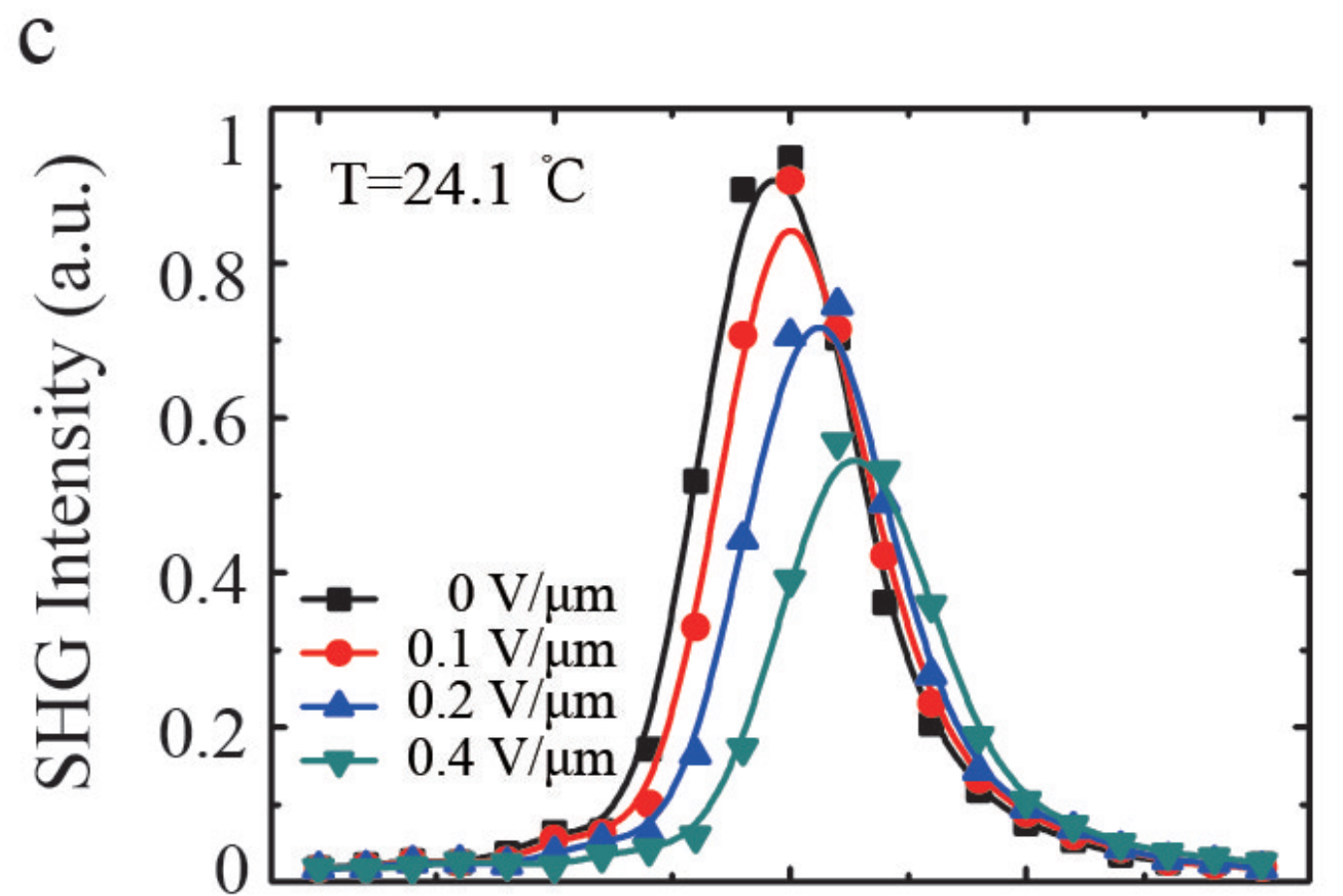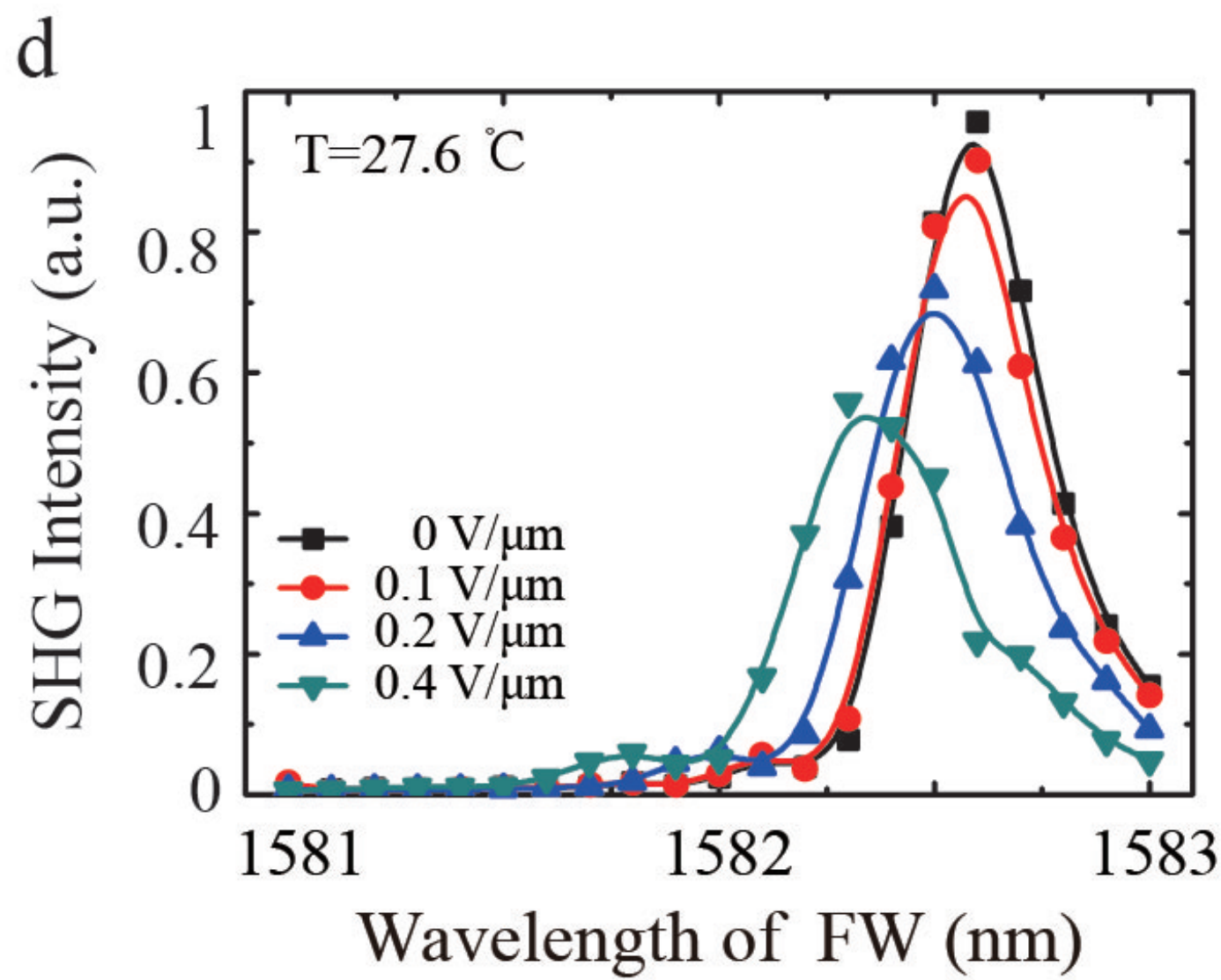

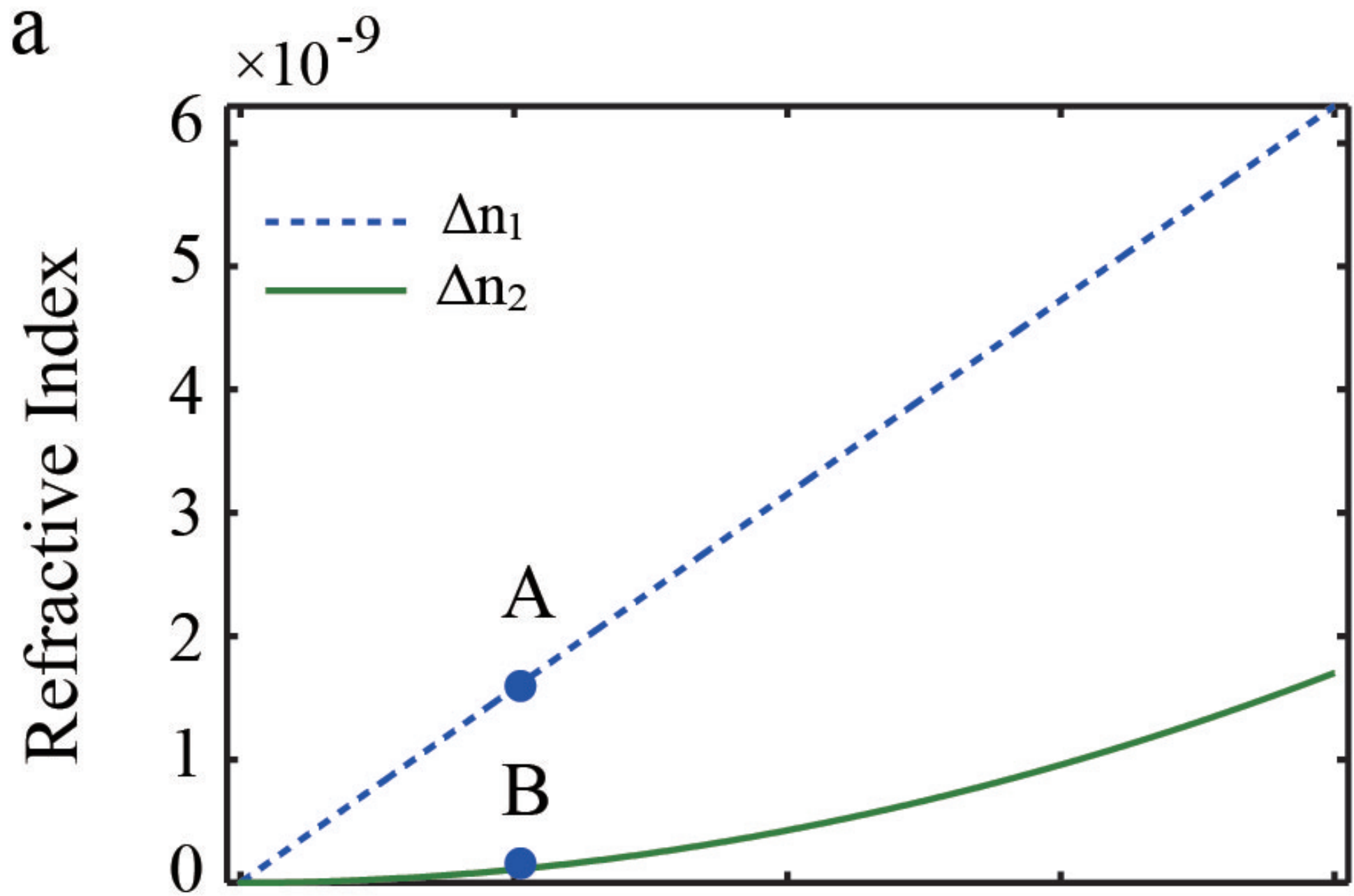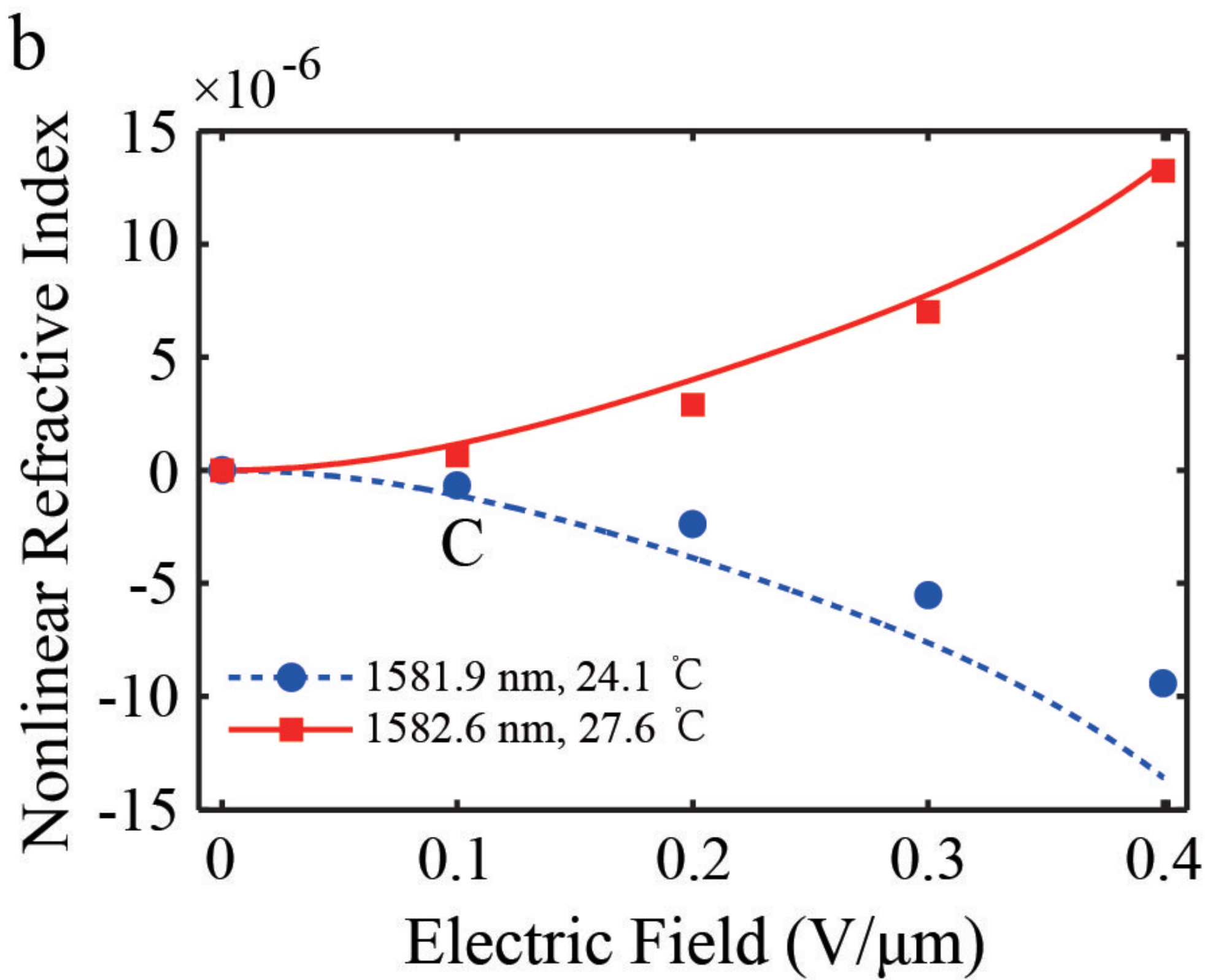

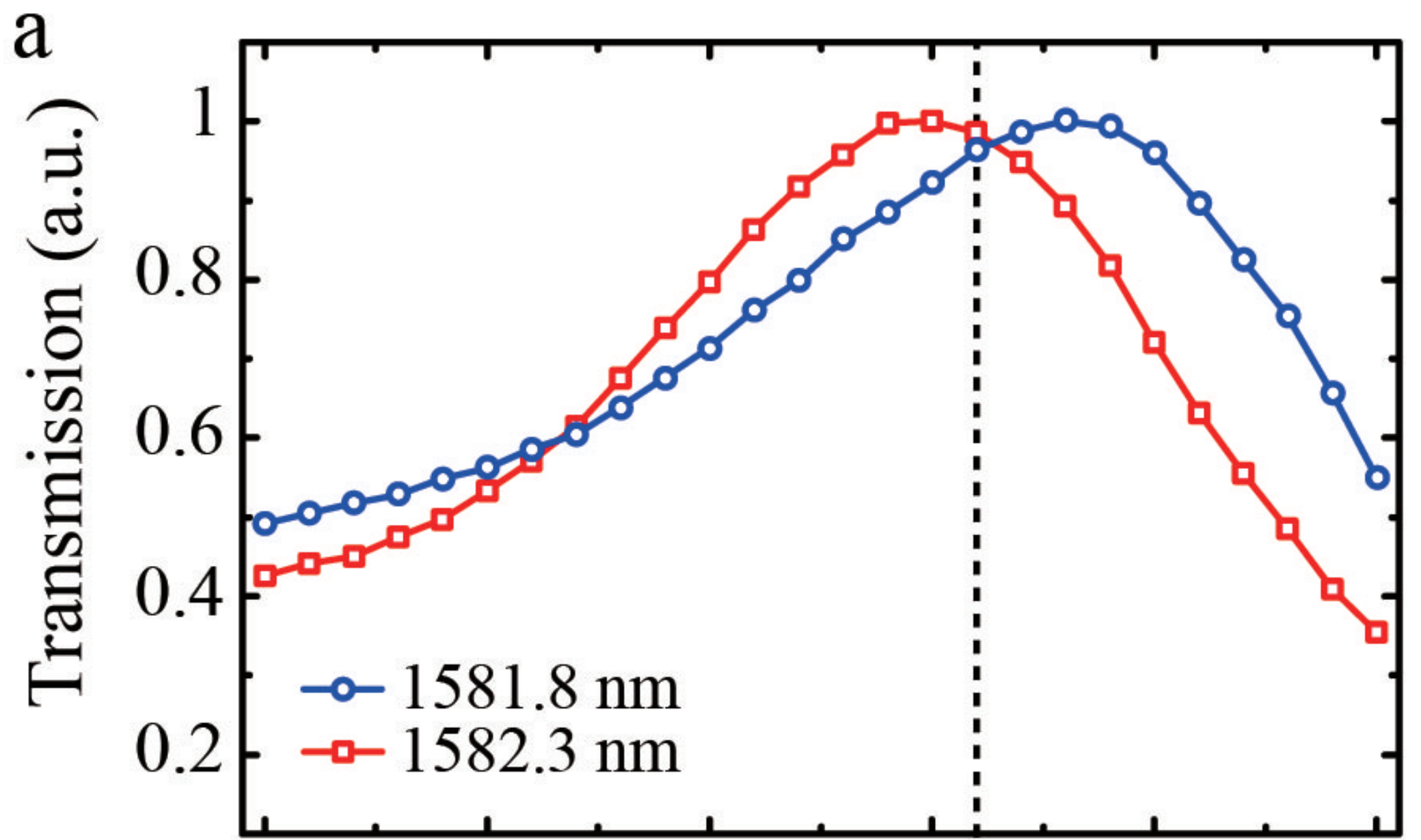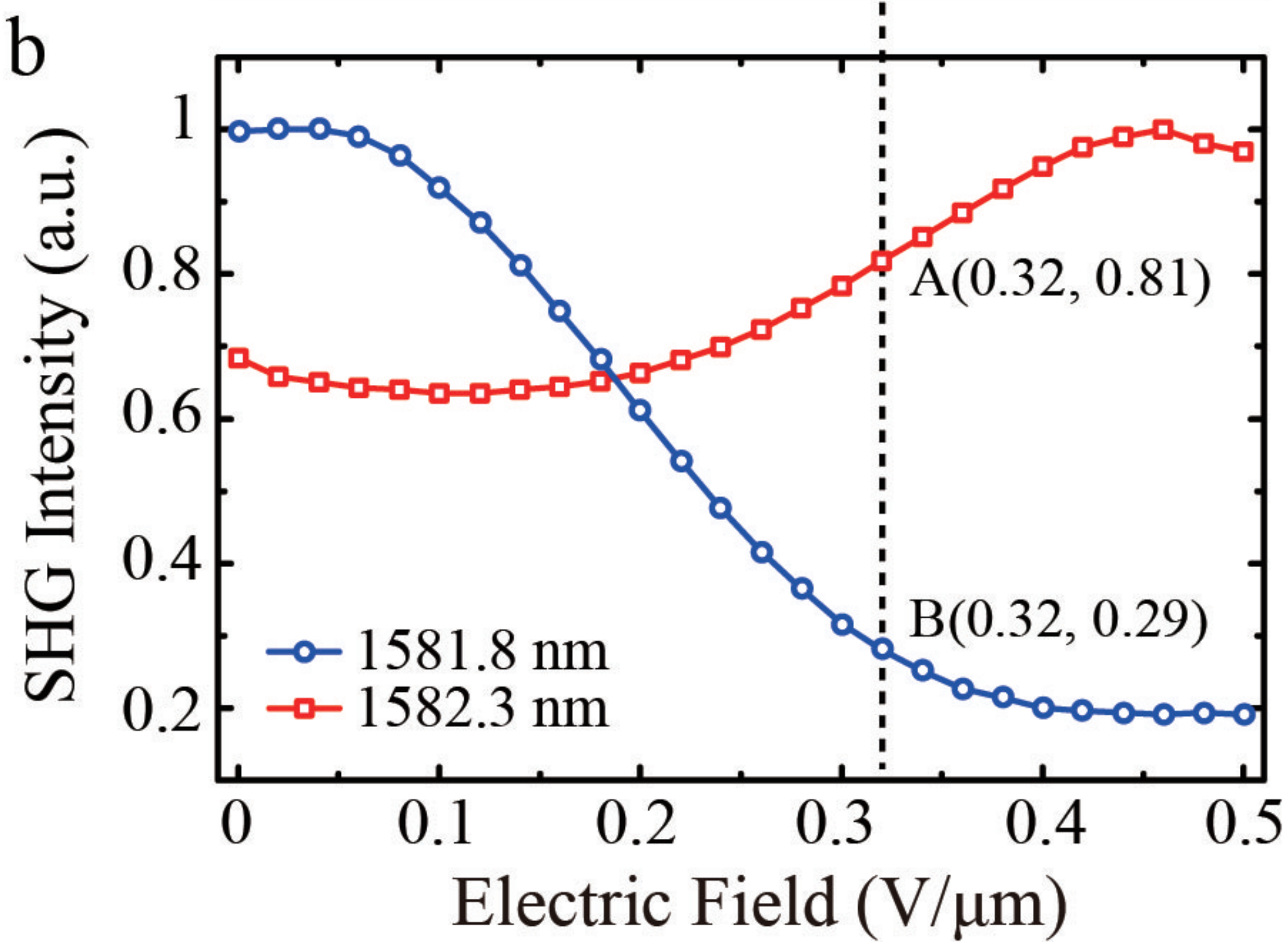

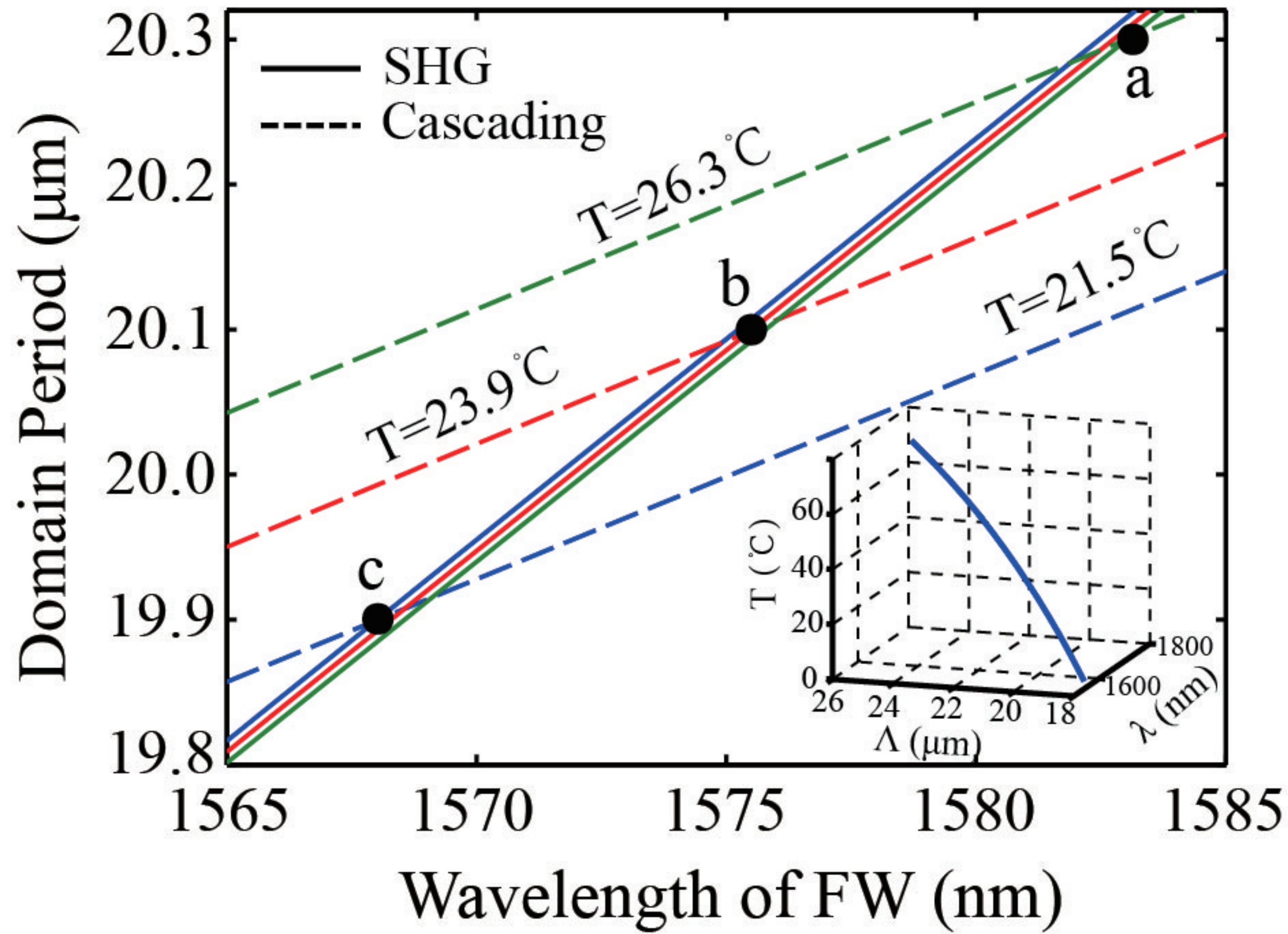